%% file: correlation_paper.tex
\documentclass[transmag]{IEEEtran}
\usepackage{latexsym}
\usepackage{graphicx}
\usepackage{amsfonts,amssymb,amsmath}
\usepackage{bm}
\usepackage{hyperref}
\def\BibTeX{{\rm B\kern-.05em{\sc i\kern-.025em b}\kern-.08em T\kern-.1667em\lower.7ex\hbox{E}\kern-.125emX}}

\usepackage{todonotes}

\newcommand{\jim}{\text{j}}

\newcommand{\vect}[1]{\bm{#1}}
\newcommand{\matr}[1]{\bm{\mathrm{#1}}}
\newcommand{\matrgreek}[1]{\bm{#1}}

\newcommand{\expect}[1]{\mathop{\mathbb{E}}\!\left\{{#1}\right\}}
\newcommand{\expectover}[2]{\mathop{\mathbb{E}}_{#2}\!\left\{{#1}\right\}}
\newcommand{\vari}[1]{\mathop{\mathbb{V}}\!\left\{{#1}\right\}}

\newcommand{\conj}[1]{{#1^\star}}
\newcommand{\herm}[1]{{#1^\text{H}}}
\newcommand{\trans}[1]{{#1^\text{T}}}

\newcommand{\abs}[1]{\left|{#1}\right|}
\newcommand{\norm}[1]{\left|\left|{#1}\right|\right|}

\usepackage[style=ieee,url=false,isbn=false,maxbibnames=6]{biblatex}
\AtBeginBibliography{\footnotesize}
\usepackage{pgfplots}
\pgfplotsset{compat=1.15}

\usetikzlibrary{arrows,shapes}

\pgfplotsset{
compat=1.11,
legend image code/.code={
\draw[mark repeat=2,mark phase=2]
plot coordinates {
(0cm,0cm)
(0.2cm,0cm)        %
(0.4cm,0cm)         %
};%
}
}

\definecolor{color0}{rgb}{0.12156862745098,0.466666666666667,0.705882352941177}
\definecolor{color1}{rgb}{1,0.498039215686275,0.0549019607843137}
\definecolor{color2}{rgb}{0.172549019607843,0.627450980392157,0.172549019607843}
\definecolor{color3}{rgb}{0.83921568627451,0.152941176470588,0.156862745098039}

\makeatletter
\let\MYcaption\@makecaption
\makeatother

\usepackage[font=footnotesize]{subcaption}

\makeatletter
\let\@makecaption\MYcaption
\makeatother

\usepackage{siunitx}

\usepackage{glossaries-extra}
\input{acronyms}

\usepackage{booktabs}

\begin{document}

\title{Statistics of the Effective Massive MIMO Channel \\ in Correlated Rician Fading}

\author{
Jens Abraham, %
Pablo Ramirez-Espinosa, %
and Torbj{\"o}rn Ekman%
\thanks{This work has been submitted to the IEEE for possible publication. Copyright may be transferred without notice, after which this version may no longer be accessible.}
\thanks{J. Abraham and T. Ekman are with the Circuits and Radio Group, Department of Electronic Systems, Faculty of Information Technology and Electrical Engineering, Norwegian University of Technology and Science, 7491 Trondheim, Norway.}
\thanks{
P. Ramirez-Espinosa is with the Department of Electronic Systems, Aalborg University, Denmark.}
}

\IEEEtitleabstractindextext{
\begin{abstract}
Massive MIMO base stations use multiple spatial diversity branches, which are often assumed to be uncorrelated in theoretical work.
Correlated branches are considered seldom since they are mathematically less tractable.
For correlated Rician fading, only the first- and second-order moments have been explored.
To describe propagation environments more accurately, full distribution functions are needed. 

This manuscript provides these functions for the maximum ratio combining effective channel, a quadratic form of a random complex normal channel vector.
Its mean vector and covariance matrix are based on a plane wave model incorporating array geometry, antenna element pattern, power angular spectra and power delay profiles.
Closed-form approximations of the distribution functions are presented, to allow the fast evaluation of many real-world scenarios. 

The statistical framework is used to show that low-directivity antenna elements provide better performance in angular constricted Rician fading with off-axis incidence than high-directivity elements.
Moreover, two base station array layouts are compared, showing that a half-circle array illuminates a cell more evenly than a uniform linear array.
With the full distribution functions available, performance can be compared over the full range of received powers and not only based on the average SNR.

\end{abstract}

}

\maketitle

\input{introduction.tex}

\input{cn_channel.tex}

\input{gqf.tex}

\input{spatial_correlation.tex}

\input{simulations.tex}

\input{discussion.tex}

\input{conclusion.tex}

\printbibliography
\end{document}

%% file: acronyms.tex
\setabbreviationstyle[acronym]{long-short}
\newacronym{3gpp}{3GPP}{3rd Generation Partnership Project}
\newacronym{aoa}{AoA}{angle of arrival}
\newacronym{aod}{AoD}{angle of departure}
\newacronym{bs}{BS}{base station}
\newacronym{cir}{CIR}{channel impulse response}
\newacronym{cn}{$\mathcal{CN}$}{complex normal}
\newacronym{cnrv}{$\mathcal{CN}$-RV}{complex normal random vector}
\newacronym{cdf}{CDF}{cumulative distribution function}
\newacronym{cgqf}{CGQF}{complex Gaussian quadratic form}
\newacronym{csi}{CSI}{channel state information}
\newacronym{ctf}{CTF}{channel transfer function}
\newacronym{dft}{DFT}{discrete Fourier transform}
\newacronym{dl}{DL}{downlink}
\newacronym{ecdf}{ECDF}{empirical cumulative distribution function}
\newacronym{epdf}{EPDF}{empirical probability density function}
\newacronym{gqf}{GQF}{Gaussian quadratic form}
\newacronym{hpbw}{HPBW}{half power beam width}
\newacronym{idft}{IDFT}{inverse discrete Fourier transform}
\newacronym{iid}{iid}{independent identically distributed}
\newacronym{iot}{IoT}{internet of things}
\newacronym{isi}{ISI}{inter-symbol-interference}
\newacronym{los}{LOS}{line of sight}
\newacronym{lpda}{LPDA}{Log-Periodic Dipole Array}
\newacronym{lsas}{LSAS}{large-scale antenna system}
\newacronym{lte}{LTE}{Long Term Evolution}
\newacronym{mgf}{MGF}{moment-generating function}
\newacronym{mimo}{MIMO}{multiple input multiple output}
\newacronym{miso}{MISO}{multiple input single output}
\newacronym{mpc}{MPC}{multi path component}
\newacronym{mrc}{MRC}{maximum ratio combining}
\newacronym{mrt}{MRT}{maximum ratio transmission}
\newacronym{mui}{MUI}{multi user interference}
\newacronym{ni}{NI}{National Instruments}
\newacronym{nlos}{NLOS}{non-line of sight}
\newacronym{ntnu}{NTNU}{Norwegian University of Technology and Science}
\newacronym{ofdm}{OFDM}{orthogonal frequency division multiplexing}
\newacronym{pas}{PAS}{power angular spectrum}
\newacronym{pdf}{PDF}{probability density function}
\newacronym{pdp}{PDP}{power delay profile}
\newacronym{reranp}{ReRaNP}{Reconfigurable Radio Network Platform}
\newacronym{rms}{rms}{root mean square}
\newacronym{rmt}{RMT}{random matrix theory}
\newacronym{rgqf}{RGQF}{real Gaussian quadratic form}
\newacronym{rv}{RV}{random variable}
\newacronym{rx}{RX}{receiver}
\newacronym{simo}{SIMO}{single input multiple output}
\newacronym{siso}{SISO}{single input single output}
\newacronym{slac}{SLAC}{stochastic local area channel}
\newacronym{snr}{SNR}{signal to noise ratio}
\newacronym{svd}{SVD}{singular value decomposition}
\newacronym{tdd}{TDD}{time division duplex}
\newacronym{tr}{TR}{time reversal}
\newacronym{tx}{TX}{transmitter}
\newacronym{ue}{UE}{user equipment}
\newacronym{ul}{UL}{uplink}
\newacronym{ula}{ULA}{uniform linear array}
\newacronym{ura}{URA}{uniform rectangular array}
\newacronym{usrp}{USRP}{Universal Software Radio Peripheral}
\newacronym{wsn}{WSN}{wireless sensor network}
\newacronym{wssus}{WSSUS}{wide-sense stationary uncorrelated scattering}

\newacronym{awgn}{AWGN}{additive white Gaussian noise}
\newacronym{sinr}{SINR}{signal-to-interference-plus-noise ratio}

%% file: introduction.tex
\section{Introduction}

\IEEEPARstart{I}{n} many massive \gls{mimo} systems, users are communicating over multiple sub-carriers with a \gls{bs} equipped with a \gls{lsas}. %
\Gls{tdd} transmission is necessary, if channel reciprocity should allow for simplified \gls{csi} acquisition.
On one hand, the radio channel is often modeled with uncorrelated antennas subject to narrow-band Rayleigh or Rician fading.
On the other hand, measurement campaigns provide evidence that the radio channel is correlated in space, time and frequency.
For massive \gls{mimo} systems, correlated Rician fading has been considered in context of the spectral efficiency \cite{ozdogan_massive_2019} and cell-free systems \cite{polegre_channel_2020,wang_uplink_2021}.
Those contexts only evaluate the first- and/or second-order behaviour of correlated narrow-band Rician fading channels.

To fully cover wide-band correlated Rician fading in \gls{lsas}, this work describes a \gls{cnrv} model with a non-trivial covariance matrix.
The vector elements are representing channel coefficients for antennas and delay taps, to allow the joint consideration of the spatial and the delay domain.
The spatial domain is parameterised by antenna element positions and corresponding \glspl{pas}, whilst the delay domain is covered by \glspl{pdp}.
The model allows to consider correlation between antennas as well as delay taps.
Hence it incorporates aspects needed to derive a physically motivated distribution for the effective channel power gain arising from \gls{mrc}.
To that extent, the influence of incidence of a deterministic channel component\footnote{The deterministic channel component can originate from a \gls{los} component or a specular component.}, its magnitude and correlation of the diffuse channel on single user performance is presented.
We provide (accurate approximate) \glspl{pdf} and \glspl{cdf} for the effective channel gain.
Moreover, the steepness of the effective channel \gls{cdf} is evaluated to investigate the local diversity \cite{abraham_local_2021} in closed-form for different outage probabilities.

Related to this work is the effective channel with selection combining for several equally correlated fading distributions given in \cite{chen_distribution_2004}.
The capacity for correlated Rayleigh \gls{mimo} channels has been derived by characterising distributions of eigenvalues of the propagation environment \cite{chiani_capacity_2003}.
For correlated Rician channels, the eigenvalue spread of the covariance matrix and the angle of the deterministic component vector with respect to the range space of the covariance matrix are key quantities of performance metrics in \cite{nabar_diversity_2005}.
Furthermore, they provide a power and a Laguerre series expansion of the effective channel.
Another power series approximation of the narrow-band \gls{mrc} effective channel \gls{snr} for antennas in a linear array is given in \cite{hon_tat_hui_performance_2005}.
The ergodic capacity of \gls{mrc} for correlated Rician channels has been evaluated in \cite{hamdi_capacity_2008}, showing that a correlated Rician fading channel can improve over an uncorrelated channel under very specific circumstances.
For uncorrelated Rician fading, massive \gls{mimo} systems have been analysed based on asymptotic expressions for the \gls{sinr} in a multi-cell system \cite{sanguinetti_theoretical_2019}.
For correlated Rician fading of a multi-cell massive \gls{mimo} scenario, different channel estimators and their resulting \gls{cnrv} parameters have been derived in \cite{ozdogan_massive_2019}.
Channel hardening and spectral efficiency in correlated Rician fading for cell-free massive \gls{mimo} has been discussed in \cite{polegre_channel_2020, wang_uplink_2021}.
A more generalised complex normal channel has been considering a Weichselberger correlation model and was analysed with focus on channel hardening and favourable propagation \cite{matthaiou_massive_2019}.
We observe, that the full statistics of correlated Rician fading channels for \gls{lsas} have not been presented so far, since they go beyond the first- and second-order statistic of the effective channel gain.

Other correlation matrix models than the Weichselberger model have been proposed in the literature.
A \emph{constant correlation matrix}, where each pair of antennas is equally correlated \cite{loyka_channel_2000}, provides a simplistic approximation.
An \emph{exponential correlation matrix}, where the difference of antenna indices determines the correlation coefficient \cite{loyka_channel_2001}, is a more realistic approximation for \glspl{ula}.
A \emph{generalised correlation matrix} model, where the correlation coefficient is a function of the euclidean distance between antenna elements, has been recently discussed in \cite{al-hussaibi_generalised_2020} and can provide approximations for more general array configurations like an \gls{ura}.
A more propagation motivated \emph{3D local scattering model} with arbitrary geometry, considering the position of array elements and the \gls{aoa} distribution of the diffuse radio channel, has been presented \cite[sec. 7.3.2]{bjornson_massive_2017}.

In this work, we add antenna element pattern explicitly to the 3D local scattering model, to analyse the effect of low or high directivity elements, that not necessarily point into the same direction.
The result is a physically meaningful parametrisation of the correlation matrix elements, whereas the aforementioned general Weichselberger model has a higher abstraction level based on eigenvalues and eigenbases.
Moreover, this manuscript extends our work on the effective channel in uncorrelated Rayleigh fading \cite{ghiaasi_effective_2019}.
\emph{Our main contribution is a method to obtain the statistics of the effective massive \gls{mimo} channel in correlated Rician fading.}
This is achieved by describing a wide-band massive \gls{mimo} system as a non-central \gls{cnrv} (sec. \ref{sec:cn_channel}).
The effective channel gain is a \gls{cgqf} for which an improved accurate approximation of the \gls{pdf} and \gls{cdf} is provided via a confluent \gls{cgqf} (sec. \ref{sec:eff_gain}).
To utilise the \gls{cnrv} model, mean and covariance are characterised considering physical properties of the propagation environment (sec. \ref{sec:correlation}):
\begin{itemize}
    \item a 3D local scattering model under consideration of \glspl{pas},
    \item antenna pattern of arbitrarily positioned and oriented \gls{bs} array elements and
    \item \glspl{pdp} with Rician fading taps.
\end{itemize} 
The findings are used to compare \glspl{bs} with a \gls{ula} or a half circle array for antenna elements with varying directivity after verification of the \gls{gqf} method with simulations (sec.~\ref{sec:simulation}).
The manuscript closes with discussion and conclusion.

\subsection{Notation}
A variable is represented as a scalar $a$, a vector $\vect{a}$ or a matrix $\matr{A}$. %
Square brackets are picking an element from a structure according to the subscript, e.g. $\left[ \vect{a} \right]_i$ is the $i$-th element of vector $\vect{a}$.
The transpose and hermitian operator are $\trans{(\cdot)}$ and $\herm{(\cdot)}$.
The expectation and variance of random variables are denoted with $\expect{\cdot}$ and $\vari{\cdot}$, respectively.
The symbols $\mathcal{CN}$, $\chi^2$ and $\chi'^2$ designate the \gls{cn}, central $\chi$-squared and non-central $\chi$-squared distribution, where e.g. $a \sim \mathcal{CN}(0,1)$ means that $a$ is distributed according to a central standard \gls{cn} distribution. 
The angles of the spherical coordinate system are $\theta$ for the azimuth ($\theta \in [0, 2 \pi]$) and $\phi$ the polar angle ($\phi \in [0, \pi]$).

%% file: cn_channel.tex
\section{Correlated Complex Normal Channel} \label{sec:cn_channel}
In this section, a correlated \gls{cnrv} $\vect{h} \sim \mathcal{CN}(\vect{\mu}, \matrgreek{\Sigma})$ is introduced to model the end-to-end propagation between a user terminal and a \gls{bs} with $M$ antenna elements observing a $N$-tap channel.
We are going to describe the mean vector $\vect{\mu}$ and covariance matrix $\matrgreek{\Sigma}$ based on antenna element locations $\vect{r}_m$ and pattern $G_m(\theta,\phi)$ for antenna element $m$ in conjunction with the \glspl{pas} $p_{l,n}(\theta,\phi)$\footnote{The term \gls{pas} is used here for both the \gls{pas} and the \gls{pdf} of the \gls{pas}, because the \gls{pdp} absorbs the power scaling.}, \glspl{pdp} $S_{l,n}$ and Rician $\mathcal{K}$-factors $\mathcal{K}_{l,n}$ of the propagation environment.
The index $l$ describes a subarray which belongs to a local area where the statistics are stationary during tap $n$ and the propagation environment is illuminated by a terminal with power $P$.
An overview of all constituents of the model is given in Fig. \ref{fig:channel_model}.
This model describes directly the observable channel coefficients and received powers at the \gls{bs} elements and highlights the impact of \gls{bs} array design on the effective channel gain.

\begin{figure}
	\centering
	\input{tikz/model_summary.tikz}
	\caption{The \gls{cnrv} channel parameters $\vect{\mu}$ and $\matrgreek{\Sigma}$ are determined by the antenna elements and physical properties of the propagation environment.
	} \label{fig:channel_model}
\end{figure}

\subsection{Propagation Environment}
The antenna element $m$, local area $l$ and delay tap $n$ will be omitted for ease of notation in the following subsections unless they are necessary to distinguish between quantities.

The \gls{pdp} coefficient $S$ describes the user transmit power that is spread into a delay tap.
The corresponding Rician factor $\mathcal{K}$ further defines the quotient between the deterministic component $\bar{h}$ and the diffuse component $\tilde{h}$ of that delay tap.

The \gls{pas} $p(\theta,\phi)$ is split into parts too, to represent the deterministic component with the incidence angles $\bar{\theta}$ and $\bar{\phi}$ as
\begin{equation}
    \bar{p}(\theta,\phi) = \delta \left(\theta - \bar{\theta} \right) \delta \left( \phi - \bar{\phi} \right)
\end{equation}
and $\tilde{p}(\theta,\phi)$ absorbing the diffuse component.
Both, $\bar{p}(\theta,\phi)$ and $\tilde{p}(\theta,\phi)$
are a \gls{pdf}.
The \gls{pas} under consideration of the $\mathcal{K}$-factor can now be described by
\begin{equation}
	p(\theta,\phi) = \frac{\mathcal{K} \bar{p}(\theta,\phi) + \tilde{p}(\theta,\phi)}{K + 1}.
\end{equation}

\subsection{Channel Realisations}
A Rician channel coefficent $h$ can be composed by superposition of a plane wave for the deterministic component $\bar{h}$ and a large number of $Z$ plane waves distributed according to the \gls{pas} to represent the diffuse component $\tilde{h}$:
\begin{equation}
	h = \bar{h} + \tilde{h}.
\end{equation}
The wavevector $\vect{k}$ of a wave with incidence angles $\theta$ and $\phi$ at wavelength $\lambda$ is given by
\begin{equation}
	\vect{k}(\theta, \phi) = \frac{2 \pi}{\lambda} \begin{bmatrix}
		\cos \theta \sin \phi \\ \sin \theta \sin \phi \\ \cos \phi
	\end{bmatrix}.
\end{equation}
This wave has phase $\varphi$ at position $\vect{r}$:
\begin{equation}
    \varphi(\theta, \phi) = \vect{k}(\theta, \phi) \cdot \vect{r}.
\end{equation}
The deterministic component $\bar{h}$ is given by the \gls{pdp} coefficient $S$, the Rician $\mathcal{K}$-factor $\mathcal{K}$, the antenna element pattern $G(\theta,\phi)$ and the phase term:
\begin{equation}
    \bar{h} = \sqrt{S \frac{K}{K + 1} G(\bar\theta, \bar\phi)} \exp \left( \jim \bar{\varphi}(\bar\theta, \bar\phi) \right). \label{eqn:deterministic_realisation}
\end{equation}
The diffuse component $\tilde{h}$ is composed with a sum over $N$ plane waves:
\begin{equation}
	\tilde{h} = \sqrt{\frac{S}{K + 1}} \sqrt{\frac{1}{N}} \sum_{z=1}^Z \sqrt{G(\tilde\theta_n,\tilde\phi_n)} a_{z} \exp \left(\jim \tilde{\varphi}_z(\tilde\theta_n,\tilde\phi_n) \right) \label{eqn:diffuse_realisation}
\end{equation}
with $a_{z}$ being \gls{iid} random magnitudes, drawn from a central standard \gls{cn} distribution.
The exponential term describes the additional phase of the incoming plane wave $n$ due to antenna element position $\vect{r}$ for an incidence wave vector $\vect{k}_{z}$.

\subsection{Mean and Auto-Covariance}
The channel realisations for all $M$ antennas and all $N$ taps are conditioned on the propagation environment and antenna array properties as described for the individual coefficient in Eqns. \eqref{eqn:deterministic_realisation} and \eqref{eqn:diffuse_realisation}.
To capture the mean and covariance of the correlated random vector, explicit mappings of indices $i$ and $j$ to antenna elements $m_i$, $m_j$ and delay taps $n_i$ and $n_j$ are necessary.
The mapping follows
\begin{align}
	i &= m_i + (n_i - 1) M \\
	j &= m_j + (n_j - 1) M 
\end{align}
to consecutively identify each antenna-tap pair uniquely.
Antenna elements $m_i$ and $m_j$ belong to the local areas $l_i$ and $l_j$, respectively.

The mean vector element $\left[\vect{\mu}\right]_i$ is simply the realisation in Eqn. \eqref{eqn:deterministic_realisation} because all variables are deterministic:
\begin{equation}
    \left[\vect{\mu}\right]_i = \expect{h_{m_i,n_i}} = \bar{h}_{m_i,n_i}.
\end{equation}

The main diagonal of the covariance matrix $\matrgreek{\Sigma}$ consists of auto-covariances of the realisations of the diffuse component in Eqn. \eqref{eqn:diffuse_realisation}:
\begin{align}
    \left[\matrgreek{\Sigma}\right]_{i,i} &= \vari{h_{m_i,n_i}} = \expect{\tilde{h}_{m_i,n_i} \conj{\tilde{h}_{m_i,n_i}}} \notag \\ &= \frac{ S}{K + 1} \iint_{\Omega} \abs{G(\theta, \phi)} \tilde{p}(\theta, \phi) d\Omega.
\end{align}
The second line of the last equation is a consequence for $Z \rightarrow \infty$ and replacement of the sum by integration over the full sphere surface $\Omega$ to account for all directions of incident waves.
Here, the pattern of antenna element $m_i$ is weighting the \gls{pas}.

\subsection{Cross-Covariances and Correlations}
The remaining entries of the covariance matrix depend on the correlation coefficient $\rho_{i,j}$ between pairs of antennas and taps:
\begin{equation}
    \left[\matrgreek{\Sigma}\right]_{i,j} = \rho_{i,j} \sqrt{\left[\matrgreek{\Sigma}\right]_{i,i} \left[\matrgreek{\Sigma}\right]_{j,j}}.
\end{equation}
The correlation coefficient $\rho_{i,j}$ is by definition:
\begin{equation}
		\rho_{i,j} = \frac{\expect{\tilde{h}_{m_i,n_i} \conj{\tilde{h}_{m_j,n_j}}}}{\sqrt{\expect{\tilde{h}_{m_i,n_i} \conj{\tilde{h}_{m_i,n_i}}} \expect{\tilde{h}_{m_j,n_j} \conj{\tilde{h}_{m_j,n_j}}}}},
\end{equation}
where only diffuse components have an influence on the covariance matrix $\matrgreek{\Sigma}$.
In the following, we will restrict our focus to radio channels exhibiting \emph{uncorrelated scattering} in the same local area ($l_i = l_j = l$)
\begin{equation}
	\rho_{i,j} = \rho_{m_i,m_j}^{t} \delta(l_i - l_j) \delta(n_i - n_j),
\end{equation}
because our interest is focused on the influence of antenna element correlations.
The correlated scattering case is left for future investigation.
It might arise from antenna elements being spaced so far from each other, that the same scatterer influeces different delay taps of those elements.
The restriction here imposes a diagonal block structure on the covariance matrix $\matrgreek{\Sigma}$, where each block $\matrgreek{\Sigma}^t$ describes the correlation between antenna elements for tap $t$.

The antenna correlation coefficient $\rho_{m_i,m_j}^{t}$ is shown in \eqref{eqn:correlation_expectation} and is an extension of the 3D local scattering model \cite[sec. 7.3.2]{bjornson_massive_2017} due to the consideration of antenna element pattern.
The local scattering model has an impact on the handling of correlation between antennas in different local areas.
We are assuming that the local areas are distant enough, such that the incoming plane waves are decorrelated.
Each incident plane wave in local area $l$ produces a direction dependent phase shift $\Delta \varphi_{m_i,m_j}$ between antenna elements at positions $\vect{r}_{m_i}$ and $\vect{r}_{m_j}$:
\begin{equation}
    \Delta \varphi_{m_i,m_j}(\theta, \phi) = \vect{k}(\theta, \phi) \cdot \left(\vect{r}_{m_i} - \vect{r}_{m_j}\right).
\end{equation}

\begin{figure*}
\normalsize
\begin{equation}
	\rho_{m_i,m_j}^n = \frac{\expectover{\sum_{z=1}^Z \sqrt{G_{m_i}(\tilde\theta_{z}, \tilde\phi_{z})} \conj{\sqrt{ G_{m_j}(\tilde\theta_{z}, \tilde\phi_{z})}} \exp \left(\jim \left(\tilde\varphi_{z,m_i}(\tilde\theta_{z}, \tilde\phi_{z}) - \tilde\varphi_{z,m_j}(\tilde\theta_{z}, \tilde\phi_{z})  \right)\right)}{\tilde\theta_{z},\tilde\phi_{z}}}{\sqrt{\expectover{\sum_{z=1}^Z \abs{G_{m_i}(\tilde\theta_{z}, \tilde\phi_{z})}}{\tilde\theta_{z},\tilde\phi_{z}} \expectover{\sum_{z=1}^Z \abs{G_{m_j}(\tilde\theta_{z}, \tilde\phi_{z})}}{\tilde\theta_{z},\tilde\phi_{z}}}} \label{eqn:correlation_expectation}
\end{equation}
\begin{equation}
    \rho_{m_i,m_j}^n = \frac{\iint_\Omega \tilde{p}_{l,n}(\theta, \phi) \sqrt{G_{m_i}(\theta, \phi)} \conj{\sqrt{G_{m_j}(\theta, \phi)}} \exp\left(\jim \Delta \varphi_{m_i,m_j}(\theta, \phi) \right) d\Omega}{\sqrt{\iint_\Omega \tilde{p}_{l,n}(\theta, \phi) \abs{G_{m_i}(\theta, \phi)} d\Omega \iint_\Omega \tilde{p}_{l,n}(\theta, \phi) \abs{G_{m_j}(\theta, \phi)} d\Omega}} \label{eqn:correlation_integral}
\end{equation}
\end{figure*}

The expectation over the sum of plane waves in Eqn. \eqref{eqn:correlation_expectation} is replaced for $Z \rightarrow \infty$ by an integration over the full sphere $\Omega$ in Eqn. \eqref{eqn:correlation_integral}.
The diffuse \glspl{pas} $\tilde{p}_{l,t}(\theta, \phi)$ conditions the incident plane waves and the integral incorporates antenna positions as well as antenna element pattern.
This allows to determine the missing elements of the correlation matrix $\matrgreek{\Sigma}$ required for a full characterisation of the \gls{cnrv} $\vect{h}$.

\subsection{Summary}
The \gls{cnrv} $\vect{h} \sim \mathcal{CN}(\vect{\mu}, \matrgreek{\Sigma})$ is fully characterised based on antenna positions, antenna pattern, \glspl{pdp}, Rician $\mathcal{K}$-factors, \glspl{pas}.
The dependencies of elements of the mean vector and the covariance matrix are summarised in the following lines, where antenna elements, delay taps and local areas are explicitly designated in the subscripts of the variables:
\begin{align}
    \left[ \vect{\mu} \right]_i \leftarrow~&S_{m_i,n_i}, \mathcal{K}_{m_i,n_i}, G_{m_i}(\theta, \phi), \bar{p}_{l_i,n_i}(\theta, \phi), \vect{r}_{m_i}, \lambda \\
    \left[ \matrgreek{\Sigma} \right]_{i,i} \leftarrow~&S_{m_i,n_i}, \mathcal{K}_{m_i,n_i}, G_{m_i}(\theta, \phi), \tilde{p}_{l_i,n_i}(\theta, \phi) \\
    \rho_{i,j} \leftarrow~&G_{m_i}(\theta, \phi), G_{m_j}(\theta, \phi), \tilde{p}_{l_i,n_i}(\theta, \phi), \tilde{p}_{l_j,n_j}(\theta, \phi), \notag \\ 
    &~\left(\vect{r}_{m_i} - \vect{r}_{m_j}\right), \lambda \\
    \left[ \matrgreek{\Sigma} \right]_{i,j} \leftarrow~&S_{m_i,n_i}, S_{m_j,n_j}, \mathcal{K}_{m_i,n_i}, \mathcal{K}_{m_j,n_j}, G_{m_i}(\theta, \phi), \notag \\
    &G_{m_j}(\theta, \phi), \tilde{p}_{l_i,n_i}(\theta, \phi), \tilde{p}_{l_j,n_j}(\theta, \phi), \left(\vect{r}_{m_i} - \vect{r}_{m_j}\right), \notag \\ &\lambda
\end{align}

The channel vector can be reshaped into a $M \times N$ matrix and multiplication of a \gls{dft} matrix from the right allows a transformation into the frequency domain if needed.
Unfortunately, the matrix form of the channel in both delay and frequency domain mixes the individual correlations between pairs of antennas and tap.

%% file: tikz/model_summary.tikz
\begin{tikzpicture}
		\node[draw, align=center, dashed, ellipse] (bslabel) {base\\station};
		\node[draw, align=center, cloud, right=1.75cm of bslabel] (channellabel) {$N$-tap\\channel};
		\node[draw, right=1.75cm of channellabel] (terminallabel) {terminal};
		\foreach \i in {1,3,5}
    	{      
            \draw[thick] (0,\i) -- (0,\i+.25);
            \draw[thick] (0,\i+.25) --++ (-0.25,.25);
            \draw[thick] (0,\i+.25) --++ (  0,.25);
            \draw[thick] (0,\i+.25) --++ ( 0.25,.25);
    	}
    	\node at (0,2.4) {\footnotesize$\vdots$};
    	\node at (0,4.4) {\footnotesize$\vdots$};
    	
    	\node at (.25,5) {\footnotesize$1$};
    	\node at (.25,3) {\footnotesize$m$};
    	\node at (.25,1) {\footnotesize$M$};
    	
    	\draw[thick, dotted] (0,3.25) plot[domain=180:0] ({0.1+4.1*sin(\x)*sin(\x)},{3.25-1.4*cos(\x)}) node[anchor=north west] {$G_m(\theta, \phi)$};
		
		\node[draw, align=center, cloud, above left=1.1cm and -2.25cm of channellabel, minimum width=5cm, minimum height=3.5cm] {};
		\node[align=center, above right=1.7cm and -.15cm of channellabel] (pas) {PAS\\$p_{l,n}(\theta,\phi)$};
		\node[align=center, left=0.1cm of pas] (rice) {Rice Factor\\$\mathcal{K}_{l,n}$};
		\node[align=center, left=0.15cm of rice] (pdp) {PDP\\$S_{l,n}$};

		\node[draw, above=2.5cm of terminallabel, thick] (k) {}; %
    	\draw[thick] (k.north) --++ (0,.25) node (ue_ant) {}; %
    	\draw[thick] (ue_ant.center) --++ (-.25,.25);
    	\draw[thick] (ue_ant.center) --++ ( .0,.25);
    	\draw[thick] (ue_ant.center) --++ (.25,.25);
    	
    	\draw (0,0) plot[domain=140:360] ({6.5+.25*sin(\x)},{3.75-.25*cos(\x)}) node[anchor=north west] {};
    	\draw (0,0) plot[domain=140:360] ({6.5+.5*sin(\x)},{3.75-.5*cos(\x)}) node[anchor=north west] {};
    	\draw (0,0) plot[domain=140:360] ({6.5+.75*sin(\x)},{3.75-.75*cos(\x)}) node[anchor=north west] {};
    	\node[below=.1cm of k,align=center] {Power\\$P$};
		\draw[decorate,decoration={brace,amplitude=10pt}] (5.8,-1.1) -- (-0.8,-1.1) node[midway,below=.4cm] {$\vect{h} \sim \mathcal{CN} \left(\vect{\mu}, \matrgreek{\Sigma}\right)$};
\end{tikzpicture}

%% file: gqf.tex
\section{Effective Channel Gain} \label{sec:eff_gain}
This section will provide the \gls{pdf} and \gls{cdf} of the effective channel gain.
This is the channel gain after combining all branches (antenna elements and taps) at the \gls{bs}.
We focus on the matched filter for the single user case, since it is the optimal result for that specific user.
Any other combination scheme under consideration of multiple users will provide poorer performance to the intended user.
To leave no user in a multi-user setting behind, interference should be suppressed by other means than interference suppressing combining.
The effective channel $\mathcal{H}$ for \gls{mrc} is:
\begin{equation}
	\mathcal{H} = \trans{\vect{w}} \vect{h} = \frac{\herm{\vect{h}} \vect{h}}{\sqrt{\norm{\vect{h}}^2_2}} = \sqrt{\herm{\vect{h}} \vect{h}}, \label{eqn:eff_channel}
\end{equation}
and the corresponding effective channel power gain $\mathcal{Q}$:
\begin{equation}
	\mathcal{Q} = \abs{\mathcal{H}}^2 = \herm{\vect{h}} \vect{h}
\end{equation}
is a \gls{cgqf} of the channel vector and a coherent summation of all vector channel elements.

Closed-form approximations of the \gls{pdf} and \gls{cdf} of \glspl{cgqf} are derived in the following.
The general idea is based on the principles of the approximation of \glspl{rgqf} \cite{ramirez-espinosa_new_real_2019}, but using the \gls{mgf} of the confluent non-central \gls{cgqf} \cite{ramirez-espinosa_new_complex_2019}.
We have reformulated the recursion in the approximation to reduce the growth rate of some auxiliary variables.
This allows to increase the approximation order, improving the accuracy of the method, enabling the analysis of \gls{lsas}.
Furthermore, the local diversity \cite{abraham_local_2021} is approximated based on the \gls{pdf} and \gls{cdf} of the confluent \glspl{cgqf}.

\subsection{Approximations of Statistics of Gaussian Quadratic Forms}

The vector $\vect{v} \sim \mathcal{CN}_N(\vect{\mu}, \matrgreek{\Sigma})$ is an $N$-element random vector, with $\vect{\mu}$ and $\matrgreek{\Sigma}$ characterising the mean vector and positive definite covariance matrix of a multivariate \gls{cn} distribution, respectively.
The vector has a quadratic form $\mathcal{Q}$ with positive semidefinite operator matrix $\matr{A}$ being:
\begin{equation}
	\mathcal{Q} = \herm{\vect{v}} \matr{A} \vect{v}.
\end{equation}
This quadratic form has the same structure as the effective channel in Eqn. \eqref{eqn:eff_channel}, where $\vect{v} = \vect{h}$ and $\matr{A} = \matr{I}$.

The vector $\vect{v}$ can be decomposed:
\begin{equation}
	\vect{v} = \matr{L} \vect{x} + \vect{\mu} = \matr{L} \left(\vect{x} + \vect{\tilde\mu} \right).
\end{equation}
such that $\vect{x} \sim \mathcal{CN}_N(0, \matr{I})$ is an \gls{iid} standard \gls{cnrv}.
The matrix $\matr{L}$ provides a mixing of the \gls{iid} variables to introduce the correlation given by $\matrgreek{\Sigma}$ (e.g. by Cholesky decomposition $\matrgreek{\Sigma} = \matr{L} \herm{\matr{L}}$) and a transformation of the mean vector $\vect{\tilde{\mu}} = \matr{L}^{-1} \vect{\mu}$.

Rewriting the quadratic form with the decomposed vector $\vect{v}$ results in:
\begin{equation}
	\mathcal{Q} = \herm{(\vect{x} + \tilde{\vect{\mu}})} \herm{\matr{L}} \matr{A} \matr{L} (\vect{x} + \tilde{\vect{\mu}})
\end{equation} 
which can be expressed in terms of eigenvalues $\lambda_i$ of $\herm{\matr{L}} \matr{A} \matr{L}$:
\begin{equation}
	\mathcal{Q} = \sum_i^{N} \lambda_i (x_i + \tilde{\mu}_i)^* (x_i + \tilde{\mu}_i) = \sum_i^N \lambda_i \abs{x_i + \tilde{\mu}_i}^2. \label{eqn:qf_eigen}
\end{equation}
This reveals the structure of a sum of $\lambda_i$ weighted non-central $\chi^2$ variables ($\abs{x_i + \tilde{\mu}_i}^2 \sim \chi^{'2}_2\left(\abs{\tilde\mu_i}^2 \right)$).
We observe that \emph{the effective channel gain of any correlated \gls{cnrv} can be rewritten as a sum of weighted independent non-central $\chi^2$ variables}, where the weights are related to the covariance matrix.

A closed-form \gls{mgf} exists for this structure \cite{ramirez-espinosa_new_complex_2019}:
\begin{equation}
	M_\mathcal{Q}(s) = \prod_{i=1}^n \exp\left( \frac{\abs{\tilde\mu_i}^2 \lambda_i s}{1-\lambda_i s} \right) \left(1 - \lambda_i s\right)^{-1},
\end{equation}
but can not be used to derive closed-forms of the corresponding \gls{pdf} and \gls{cdf}. 
Nonetheless, using a slightly modified \gls{mgf} which converges for approximation order $m \rightarrow \infty$ to the intended \gls{cgqf} \cite{ramirez-espinosa_new_real_2019} gives the following approximation for the \gls{pdf}:
\begin{equation}
	f_\mathcal{Q}(x) \approx M_\mathcal{Q}\left(\frac{1-m}{x}\right) \frac{(m-1)^m}{x^{m+1} (m-1)!} U_m\left(\frac{1-m}{x} \right), \label{eqn:pdf_original}
\end{equation}
as well as \gls{cdf}
\begin{equation}
	F_\mathcal{Q}(x) \approx M_\mathcal{Q}\left(\frac{1-m}{x}\right) \sum_{k=0}^{m-1} \frac{(m-1)^k}{x^{k} k!} U_k\left(\frac{1-m}{x}\right), \label{eqn:cdf_original}
\end{equation}
with auxiliary variables:
\begin{align}
	U_k(s) &= \sum_{j=0}^{k-1} \binom{k-1}{j} V_{k-1-j}(s) U_j(s) \label{eqn:Dk_original} \\
	V_t(s) &= t! \sum_{i=1}^{n} \lambda_i^{t+1} \frac{(t+1) \abs{\tilde\mu_i}^2 - \lambda_i s + 1}{\left(1 - \lambda_i s \right)^{t+2}}. \label{eqn:gt_original}
\end{align}

The variable $U_k(s)$ can be calculated by recursion and builds on $g_t(s)$.
Both auxiliary variables grow fast and overflow a floating point number, when the approximation order $m$ grows large.
Detection of the overflow allows to gracefully identify the maximum approximation order, where results are still valid.

To improve the numerical properties of the approximation, we reformulate the auxiliary variables in \eqref{eqn:Dk_original} and \eqref{eqn:gt_original}.
The growth rate can be reduced by redistributing the fast growing faculty terms $k!$ and $t!$ as follows:
\begin{align}
	\tilde{U}_k(s) &= (-s)^k \frac{U_k(s)}{k!} \\
	\tilde{V}_t(s) &= (-s)^t \frac{V_{t-1}(s)}{(t-1)!}.
\end{align}
This eliminates the binomial in \eqref{eqn:Dk_original} and gives the modified auxiliary variables:
\begin{align}
	\tilde{U}_k(s) &= \frac{1}{k} \sum_{j=0}^{k-1} \tilde{V}_{k-j}(s) \tilde{U}_j(s), \label{eqn:Dk_mod} \\
	\tilde{V}_t(s) &= (-s)^t \sum_i^n \lambda_i^t  \frac{t \abs{\tilde\mu_i}^2 - \lambda_i s + 1}{\left(1 - \lambda_i s\right)^{t+1}}.
\end{align}
Additionally, the approximations of the \gls{pdf} and \gls{cdf} are simplified to:
\begin{align}
	f_\mathcal{Q}(x) &\approx \frac{m}{x} M_Q \left(\frac{1-m}{x}\right) \tilde{U}_m\left(\frac{1-m}{x}\right), \label{eqn:pdf_mod} \\
	F_\mathcal{Q}(x) &\approx M_Q \left(\frac{1-m}{x}\right) \sum_{k=0}^{m-1} \tilde{U}_k\left(\frac{1-m}{x}\right). \label{eqn:cdf_mod}
\end{align}
and the local diversity $\mathcal{D}$ \cite{abraham_local_2021} of a quadratic form follows:
\begin{equation}
	\mathcal{D}(x) = x \frac{f_\mathcal{Q}(x)}{F_\mathcal{Q}(x)} \approx m \frac{\tilde{U}_m(\frac{1-m}{x})}{\sum_{k=0}^{m-1} \tilde{U}_k(\frac{1-m}{x})}.
\end{equation}
This form allows to calculate the local diversity as a byproduct of the \gls{pdf} calculation.

%% file: spatial_correlation.tex
\section{Examples of Spatial Correlation} \label{sec:correlation}
This section demonstrates the effect of different \gls{pas}, as well as antenna pattern, on the spatial correlation between antenna elements.
We will show the differences to the classic case of isotropic antennas in Rayleigh fading.

The spatial correlation is completely described by the covariance matrix $\matrgreek{\Sigma}$ and takes into account both, the \gls{pas} of the diffuse channel and the directivity of the antenna elements.
Here, we only consider uncorrelated scattering for simplicity, but scenarios with correlation between multiple taps can be covered too.
First, different \gls{pas} (see Fig. \ref{fig:pas}) for \glspl{ula} with omni-directional antenna elements are presented and their influence on the correlation coefficients is analysed.
Then, antenna element patterns are added into the consideration.

\subsection{Power Angular Spectra}

\subsubsection{Omni-directional Channels}
The classic Rayleigh channel is omni-directional in two dimensions and has a uniform \gls{pas} in azimuth restricted to a single polar angle of $\phi = \pi / 2$: 
\begin{equation}
    p(\theta, \phi) = \tilde{p}(\theta, \phi) =  \frac{1}{2\pi} \delta(\phi - \frac{\pi}{2}).
\end{equation}
However, a Rayleigh fading envelope does not necessarily require the diffuse component to be omni-directional.
Evaluating the correlation coefficient for a uniformly spaced $x$-oriented \gls{ula} with $\Delta d$ element spacing gives:
\begin{align}
	\rho_{i,j}^\text{omni} &= \frac{\delta(n_i - n_j)}{2 \pi} \int_0^{2 \pi} \exp \left(\jim 2 \pi (m_i - m_j) \frac{\Delta d}{\lambda} \cos \theta \right) d\theta \\
	&= \delta(n_i - n_j) J_0\left(2 \pi (m_i - m_j) \frac{\Delta d}{\lambda}\right),
\end{align}
where the $\delta$-function ensures uncorrelated taps and $J_0(\cdot)$ is the zero-order Bessel function of the first kind.

The Rice channel needs to account for the $\mathcal{K}$-factor in addition, but the correlation properties stay the same and $\rho_{i,j}^\text{omni}$ provides the correlation coefficient.
The corresponding \gls{pas}, where the available channel power has been normalised, has the following \gls{pdf} \cite[Sec. 6.4.4]{durgin_space-time_2003}:
\begin{equation}
    p(\theta,\phi) = \frac{1}{2 \pi (K+1)} \left[1 + 2 \pi K \delta(\theta-\bar\theta) \right] \delta(\phi - \frac{\pi}{2}),
\end{equation}
where the angle $\bar\theta$ defines the direction of the incoming wave responsible for the deterministic part of the channel.

\subsubsection{Sector Channel}
It is more common in outdoor propagation scenarios with elevated \gls{bs}, that the diffuse part is restricted to a sector with an opening angle $\psi$,
We are continuing with two-dimensional propagation coming from the horizon and define the sector \gls{pas} as:
\begin{equation}
	\tilde{p}(\theta,\phi) = \delta(\phi - \frac{\pi}{2}) \begin{cases}
		\frac{1}{\psi} & \text{with }\tilde\theta - \frac{\psi}{2} \le \theta \le \tilde\theta + \frac{\psi}{2} \\ 0 & \text{otherwise}
	\end{cases}
\end{equation}
where $\tilde\theta$ is the directional centre of the diffuse part.

The correlation coefficient resulting from propagation from a uniform diffuse sector impinging on a \gls{ula} is
\begin{align}
	\rho_{i,j}^\text{uni} = \frac{\delta(n_i - n_j)}{\psi} \int_{\tilde\theta - \frac{\psi}{2}}^{\tilde\theta + \frac{\psi}{2}} \exp \left( \jim 2 \pi (m_i - m_j) \frac{\Delta d}{\lambda} \cos \theta \right) d\theta.
\end{align}
This integral can be solved numerically.

A tapered sector model can be achieved with a von Mises distribution\footnote{The circular equivalent to the real Gaussian distribution.}, to avoid discontinuities in the \gls{pas}.
The \gls{pdf} for the diffuse channel \gls{pas} is:
\begin{equation}
	\tilde{p}(\theta,\phi) = \delta(\phi) \frac{\exp \left(\kappa \cos \left(\theta - \tilde\theta\right)\right)}{2 \pi I_0(\kappa)}.
\end{equation}
The correlation coefficient can be calculated by numerically solving \eqref{eqn:correlation_integral}.

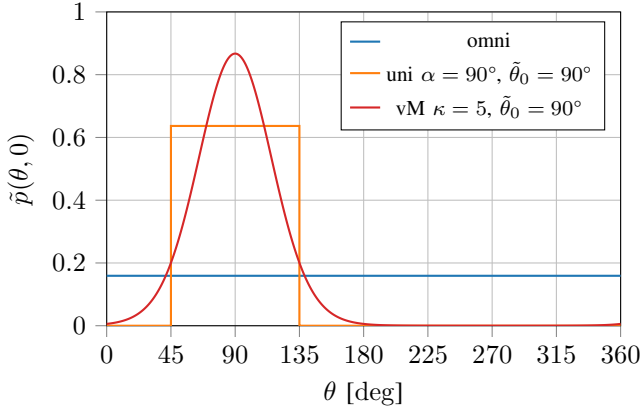
\begin{figure}
	\input{tikz/pas.tikz}
	\caption{\glspl{pas} for three different scenarios are shown.} \label{fig:pas}
\end{figure}

\begin{figure}
	\input{tikz/rho_vs_pas.tikz}
	\caption{The correlation between two omni-directional elements spaced at a distance in $x$-direction is shown for different \glspl{pas}.} \label{fig:element_correlation}
\end{figure}
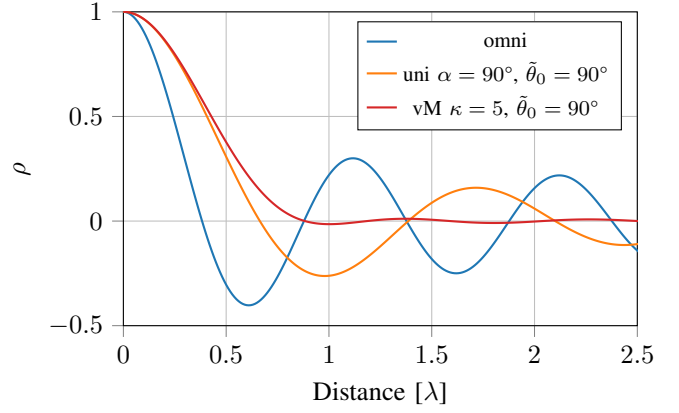

In Fig. \ref{fig:pas} three different diffuse \gls{pas} are displayed.
The corresponding correlation function for two omni-directional elements separated in $x$-direction are shown in Fig. \ref{fig:element_correlation}.
It is clear that the actual distribution of the \gls{pas} has a strong influence on the spatial correlation between antenna elements.
The wider the sector of the diffuse component is, the slower is the decay of the correlation coefficient magnitude.

\subsection{Antenna Element Pattern}
The actually observable diffuse part of the channel depends not only on the direction of the incoming diffuse waves, but on the antenna element pattern in addition.
For simplicity, we are continuing with the specialisation to two-dimensions.

The directivity $D$ of each antenna element is described by the shape of the antenna pattern $F(\theta, \phi)$ and the maximum directivity $D_0$:
\begin{equation}
	D(\theta, \phi) = D_0 F(\theta, \phi).
\end{equation}
A generic uni-directional two-dimensional antenna pattern can be modeled as a cosine to the power of $\zeta$ \cite{balanis_antenna_2016}, where $\theta_0$ fixes the azimuth angle for $D_0$: 
\begin{equation}
	F(\theta, \phi) = \begin{cases}
		\cos^\zeta(\theta - \theta_0) \delta(\phi - \frac{\pi}{2}) & \theta_0 - \frac{\pi}{2} \le \theta \le \theta_0 + \frac{\pi}{2} \\
		0 & \text{elsewhere}.
	\end{cases}
\end{equation}
The higher $\zeta$ is, the more directional the antenna pattern.
The \gls{hpbw} can be derived by evaluating $F(\theta, \phi) = 1/2$:
\begin{equation}
	\theta_\text{HPBW} = 2 \arccos\left(\sqrt[\zeta]{ \frac{1}{2}}\right). 
\end{equation}
The maximum directivity with respect to an isotropic source can be evaluated by:
\begin{equation}
	D_0 = \frac{4 \pi}{\iint_\Omega F(\theta, \phi)d\Omega}.
\end{equation}

\begin{figure}
    \centering
    \input{tikz/antenna_pattern.tikz}
    \caption{Directivity of the uni-directional generic $\cos^\zeta(\theta - \theta_0)$ antenna element for $\theta_0 = \ang{90}$}
    \label{fig:ant_pattern}
\end{figure}
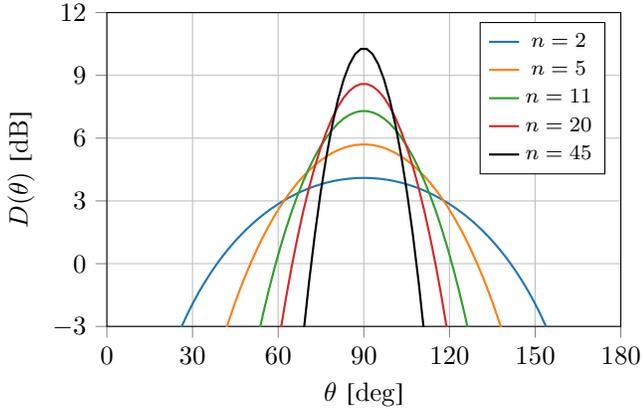

The pattern for different $\zeta$ is shown in Fig. \ref{fig:ant_pattern} and the maximum directivity and \gls{hpbw} are presented in Table \ref{tab:antenna_pattern}.

\begin{table}
\centering
\caption{Properties of the generic uni-directional $\cos^\zeta(\theta-\theta_0)$ antenna element} \label{tab:antenna_pattern}
	\begin{tabular}{lcccccc}
		\toprule
			$\zeta$ & 0 & 2 & 5 & 11 & 20 & 45 \\
		\cmidrule(lr){2-7}
			$D_0~[\si{\deci\bel}]$ & 1.0 & 4.1 & 5.7 & 7.3 & 8.6 & 10.3 \\
			$\theta_\text{HPBW}$ & (\ang{180}) & \ang{90} & \ang{59} & \ang{40} & \ang{30} & \ang{20} \\
		\bottomrule
	\end{tabular}
\end{table}

The antenna gain in section \ref{sec:cn_channel} is connected to the directivity via the antenna efficiency $\epsilon$:
\begin{equation}
	G(\theta,\phi) = \eta D(\theta,\phi)
\end{equation}

\begin{figure}
	\input{tikz/rho_vs_dir.tikz}
	\caption{The correlation between two antenna elements spaced at a distance of $\num{0.5}~\lambda$ is shown for different squint angles.} 	\label{fig:directionality_correlation}
\end{figure}
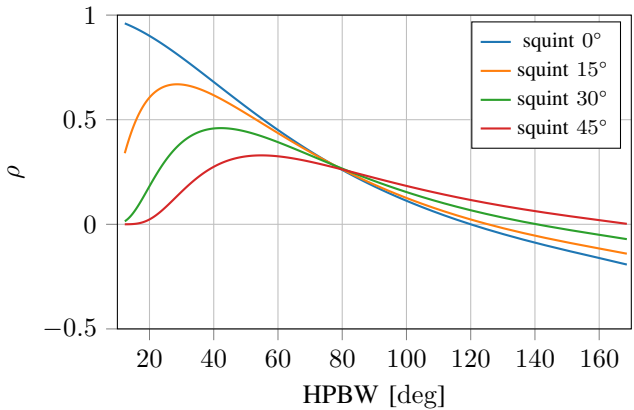

For the generic $cos^\zeta(\theta-\theta_0)$, the correlation coefficient in Eqn. \eqref{eqn:correlation_integral} can be simplified as shown in Eqn. \eqref{eqn:gen_cos_corr_coeff}.
\begin{figure*}
\begin{equation}
    \rho_{m_i,m_j}^t = \frac{D_0}{4\pi} \int_{\max\left(\theta_{0_i}-\frac{\pi}{2},\theta_{0_j}-\frac{\pi}{2}\right)}^{\min\left(\theta_{0_i}+\frac{\pi}{2},\theta_{0_j}+\frac{\pi}{2}\right)} \sqrt{\cos^\zeta(\theta - \theta_{0_i}) \cos^\zeta(\theta - \theta_{0_j})} \exp\left(\jim \Delta \varphi_{m_i,m_j}(\theta, \frac{\pi}{2}) \right) d\theta \label{eqn:gen_cos_corr_coeff}
\end{equation}
\end{figure*}
Fig. \ref{fig:directionality_correlation} depicts the influence of the directivity of the antenna elements on the correlation coefficient at an element spacing of $\lambda/2$.
The more directive the antenna elements are, the higher the correlation coefficient for \glspl{ula} (squint angle \ang{0}).
If the elements are squinting into different directions (e.g. if they are distributed over an arc), then the behaviour changes and the correlation falls off once both beams stop to overlap.

%% file: tikz/pas.tikz
\begin{tikzpicture}
\begin{axis}[
legend pos=north east,
legend columns=1, 
tick align=outside,
tick pos=left,
xlabel={$\theta$ [$\deg$]},
xmajorgrids,
xmin=0, xmax=360,
xtick={0,45,90,135,180, 225, 270, 315, 360},
ylabel={$\tilde{p}(\theta, 0)$},
ymajorgrids,
ymin=0, ymax=1,
width=.95\linewidth, height=.65\linewidth,
]
\addplot [thick, color0] plot coordinates {(0,0.159154943) (360,0.159154943)};
\addlegendentry{\footnotesize omni};
\addplot [thick, color1] plot coordinates {(0,0) (45,0) (45,0.636619772) (135,0.636619772) (135,0) (360,0)};
\addlegendentry{\footnotesize uni $\alpha = \ang{90}$, $\tilde\theta_0 = \ang{90}$};
\addplot [thick, color3] table[x index=0, y index=1 ] {data/pas_sec_vM_ula_5-90.csv};
\addlegendentry{\footnotesize vM $\kappa = 5$, $\tilde\theta_0 = \ang{90}$};
\end{axis}
\end{tikzpicture}

%% file: tikz/rho_vs_pas.tikz
\begin{tikzpicture}
\begin{axis}[
legend pos=north east,
legend columns=1, 
tick align=outside,
tick pos=left,
xlabel={Distance [$\lambda$]},
xmajorgrids,
xmin=0, xmax=2.5,
ylabel={$\rho$},
ymajorgrids,
ymin=-.5, ymax=1,
width=.95\linewidth, height=.65\linewidth,
]
\addplot [thick, color0] table[x index=0, y expr=\thisrowno{1} ] {data/rho_omni_ula.csv};
\addlegendentry{\footnotesize omni};
\addplot [thick, color1] table[x index=0, y expr=\thisrowno{1} ] {data/rho_sec_ula_90-90.csv};
\addlegendentry{\footnotesize uni $\alpha = \ang{90}$, $\tilde\theta_0 = \ang{90}$};
\addplot [thick, color3] table[x index=0, y expr=\thisrowno{1} ] {data/rho_sec_vM_ula_5-90.csv};
\addlegendentry{\footnotesize vM $\kappa = 5$, $\tilde\theta_0 = \ang{90}$};
\end{axis}
\end{tikzpicture}

%% file: tikz/antenna_pattern.tikz
\begin{tikzpicture}
\begin{axis}[
legend pos=north east,
legend columns=1, 
tick align=outside,
tick pos=left,
xlabel={$\theta$ [$\deg$]},
xmajorgrids,
xmin=0, xmax=180,
xtick={0,30,60,90,120,150,180},
ylabel={$D(\theta)$ [\si{\deci\bel}]},
ymajorgrids,
ymin=-3, ymax=12,
ytick={-3,0,3,6,9,12},
width=.95\linewidth, height=.65\linewidth,
domain=0:180,
]
\addplot [thick,samples=90,color0] {4.1 + 10*log10(cos(x-90)^2)};
\addlegendentry{\footnotesize $n=2$};
\addplot [thick,samples=90,color1] {5.7 + 10*log10(cos(x-90)^5)};
\addlegendentry{\footnotesize $n=5$};
\addplot [thick,samples=90,color2] {7.3 + 10*log10(cos(x-90)^11)};
\addlegendentry{\footnotesize $n=11$};
\addplot [thick,samples=90,color3] {8.6 + 10*log10(cos(x-90)^20)};
\addlegendentry{\footnotesize $n=20$};
\addplot [thick,samples=90,black] {10.3 + 10*log10(cos(x-90)^45)};
\addlegendentry{\footnotesize $n=45$};
\end{axis}
\end{tikzpicture}

%% file: tikz/rho_vs_dir.tikz
\begin{tikzpicture}
\begin{axis}[
tick align=outside,
tick pos=left,
xlabel={HPBW [$\deg$]},
xmajorgrids,
xmin=10, xmax=170,
ylabel={$\rho$},
ymajorgrids,
ymin=-.5, ymax=1,
width=.95\linewidth, height=.65\linewidth,
]

\addplot [thick, color0] table[x index=1, y expr=(\thisrowno{2}) ] {data/rho_n_squint_0.0.csv};
\addlegendentry{\footnotesize squint \ang{0}};

\addplot [thick, color1] table[x index=1, y expr=(\thisrowno{2}) ] {data/rho_n_squint_15.0.csv};
\addlegendentry{\footnotesize squint \ang{15}};

\addplot [thick, color2] table[x index=1, y expr=(\thisrowno{2}) ] {data/rho_n_squint_30.0.csv};
\addlegendentry{\footnotesize squint \ang{30}};

\addplot [thick, color3] table[x index=1, y expr=(\thisrowno{2}) ] {data/rho_n_squint_45.0.csv};
\addlegendentry{\footnotesize squint \ang{45}};

\end{axis}
\end{tikzpicture}

%% file: simulations.tex
\section{Simulations} \label{sec:simulation}
In the following section, simulation results are presented, to verify the approximations due to the confluent \gls{cgqf}.
To illustrate the versatility of the method, we compare a \gls{ula} \gls{bs} with a half-circle \gls{bs} layout for antenna elements with varying directivity.
This shall demonstrate how the introduced model allows a simple performance evaluation of correlated Rician radio channels under consideration of antenna element pattern, array layout and Rician fading channels. 

\subsection{Verification}

To verify that the analytic approximations provide accurate results, a number of diffuse plane waves impinging on an \gls{ula} with isotropic antenna elements are simulated. 
The 32 antenna element \gls{ula} with $\lambda/2$ spacing in $x$-direction is situated in a single tap Rician fading environment with $\mathcal{K}$-factor 4 for four different \gls{pas}:
\begin{itemize}
	\item uncorrelated
	\item omnidirectional diffuse scattering
	\item von Mises scattering with $\mathcal{K}=5$ aligned with deterministic component ( $\bar{\theta}=\tilde{\theta_0}=\ang{70}$)
	\item von Mises scattering with $\mathcal{K}=5$ squinting with respect to the deterministic component ($\bar{\theta}=\ang{70}, \tilde{\theta_0} = \ang{90}$)
\end{itemize}
The uncorrelated simulation adhering to Eqns. \eqref{eqn:deterministic_realisation} and \eqref{eqn:diffuse_realisation} is generating independent plane waves for each antenna element, whilst the other three simulations have the same plane waves impinging on all elements.
The \glspl{ecdf} are based on \num{1e6} trials with 800 plane waves forming the diffuse channel according to the \gls{pas}.

These verification settings cover wide-band results up to a certain number of taps too, since both taps and antennas link to \gls{cnrv} elements. The multi-antenna single-tap case allows for a fully populated correlation matrix in our uncorrelated scattering setting and is therefore more challenging than a single antenna 32 tap scenario.

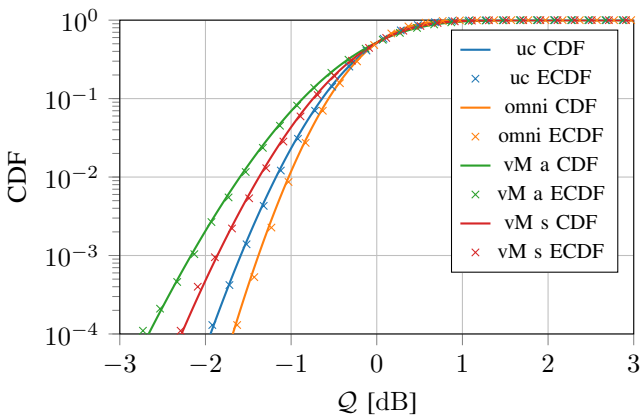
\begin{figure}
	\input{tikz/verification_cdf.tikz}
	\caption{Verification of the \glspl{cdf} approximated by \glspl{cgqf} for a 32 antenna element \gls{ula} in a single tap fading environment with $\mathcal{K}$-factor 4. The \glspl{ecdf} are generated from simulation results. Four different \gls{pas} are evaluated: uncorrelated (uc), omni-directional (omni), von Mises aligned (vM a) and von Mises squinting (s). The approximations provide accurate results for the correlated and uncorrelated cases.} \label{fig:verification_cdf}
\end{figure}

The \glspl{cdf} are shown in Fig. \ref{fig:verification_cdf}.
Simulations and approximations give consistent results for all cases.
It is important to note, that uncorrelated antenna elements do \emph{not} in general provide the best results once the channel shows Rician fading.
It depends on the superposition of the phases that the deterministic and diffuse component cause on the antenna elements.
A certain correlation between close antenna elements improves the situation if the diffuse component is orthogonal to the deterministic component.
This reduces the probability that the diffuse component acts destructively on the deterministic component, since they align over close antenna elements.
This effect is visible both, in comparison with the uncorrelated and omni-directional \gls{pas} and for the aligned and squinting von Mises scattering.
Nonetheless, for real scenarios it is highly probable the the deterministic and the diffuse component are aligned (e.g. von Mises aligned case), leading to a loss of performance over the uncorrelated case.

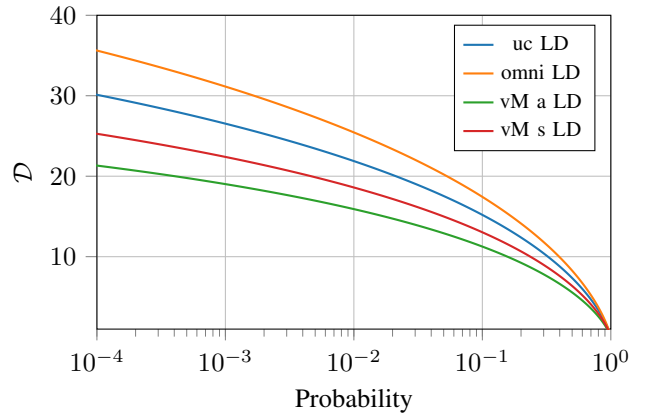
\begin{figure}
	\input{tikz/verification_ld.tikz}
	\caption{Local diversity approximated by \glspl{cgqf} for a 32 antenna element \gls{ula} in a single tap fading environment with $\mathcal{K}$-factor 4 and isotropic antenna pattern.} \label{fig:verification_ld}
\end{figure}

The corresponding local diversity is presented in Fig. \ref{fig:verification_ld}.
The classic diversity would be 32 for a \gls{bs} with 32 uncorrelated antenna elements.
The local diversity at interesting outage probabilities is heavily depending on the \gls{pas} of the diffuse component.
Additionally, it is not predictable from the number of array elements only \cite{abraham_local_2021}.

\subsection{Array Configurations}
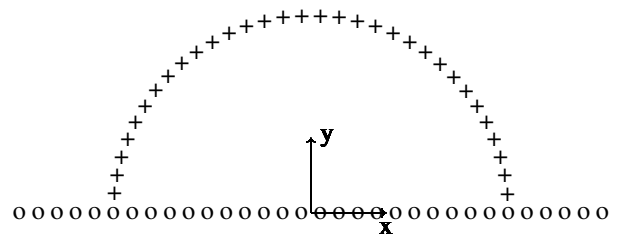
\begin{figure}
\centering
\begin{tikzpicture}
	\foreach \i in {1,...,32} {
		\node at ({5.25*cos(\i * 5.45)/2}, {5.25*sin(\i * 5.45)/2}) {+}; %
		\node at ({(\i-16.5)*0.5/2}, 0) {o};
		\draw[->] (0,0) -- (1,0) node[anchor=north] {x};
		\draw[->] (0,0) -- (0,1) node[anchor=west] {y};
	}
\end{tikzpicture}
\caption{Array configuration of the \gls{ula} (o) and half circle array (+).} \label{fig:array_config}
\end{figure}

Here, we want to analyse the implications of two different \gls{bs} array configurations in aligned von Mises scattering from different angles with $\mathcal{K}$-factors zero and 4.
An \gls{ula} is the reference configuration, since it is used abundantly for theoretical discussions due to its mathematical tractability.
Equipped with directional antenna elements it could model a \gls{bs} mounted to a building edge fairly well. 
As alternative, the elements will be distributed over a half circle to reduce the extent in one direction, whilst increasing it into the other.
Hence, with the same directional elements, a wider sector will be illuminated more evenly, reducing the maximum gain into broadside direction. 
Both configurations are sketched in Fig.~\ref{fig:array_config}.

\begin{figure*}
\begin{subfigure}[b]{.5\linewidth}
\centering \input{tikz/bs_comp_0_60.tikz}
\caption{$\mathcal{K}=0, \tilde\theta_0 = \ang{60}$}
\end{subfigure}%
\begin{subfigure}[b]{.5\linewidth}
\centering \input{tikz/bs_comp_4_60.tikz}
\caption{$\mathcal{K}=4, \bar\theta = \tilde\theta_0 = \ang{60}$}
\end{subfigure}
\begin{subfigure}[b]{.5\linewidth}
\centering \input{tikz/bs_comp_0_30.tikz}
\caption{$\mathcal{K}=0, \tilde\theta_0 = \ang{30}$}
\end{subfigure}%
\begin{subfigure}[b]{.5\linewidth}
\centering \input{tikz/bs_comp_4_30.tikz}
\caption{$\mathcal{K}=4, \bar\theta = \tilde\theta_0 = \ang{30}$}
\end{subfigure}%
\caption{The plots show the effective channel for different directivities (colours), \glspl{pas} with $\mathcal{K} = 5$ of a \gls{ula} (solid lines), and half circle array (dashed lines). The \gls{ula} performance is more dependent on broadside incidence of the waves, than the half circle array, especially for higher directivity elements. In all shown cases is lower directivity more beneficial and the half circle arrangement trades peak directivity towards broadside with increased directivity towards the end-fire direction of the \gls{ula}.} \label{fig:bs_comparison}
\end{figure*}
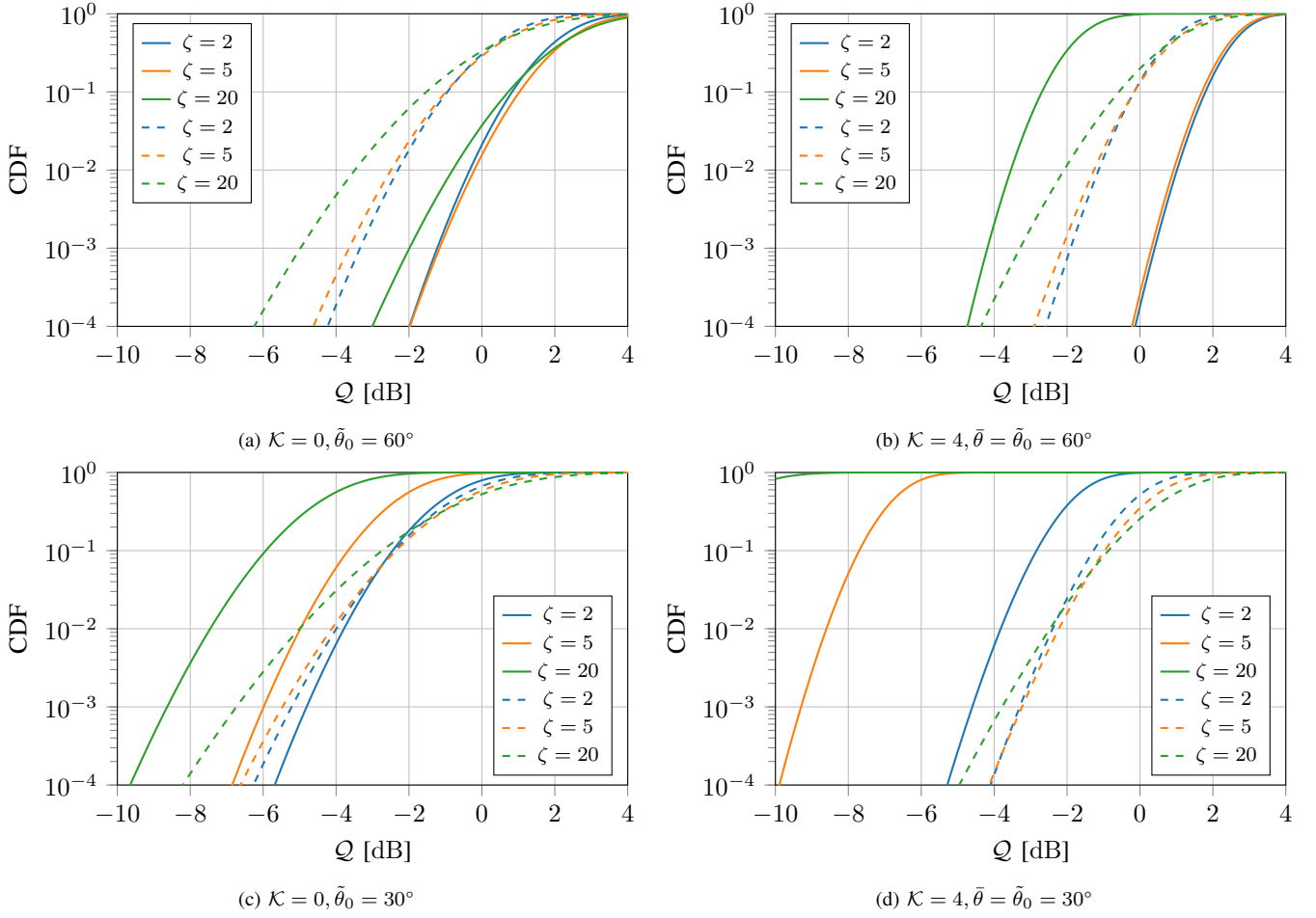

Fig. \ref{fig:bs_comparison} shows results for different von Mises \glspl{pas}.
Solid lines and dashed lines indicate the \gls{ula} and half circle \gls{bs} layout, respectively.
In general, high directivity of the antenna elements is not giving advantages in the considered cases.
The reason is that the main direction of the elements has to match the direction of the incoming waves, but it is not steerable.
The penalty for higher gain elements is lower in the half circle arrangement, since at least a few antennas are pointing towards a possible user.

For the low directivity elements, a deterministic channel component can be used efficiently, since the array factor allows coherent combination.
In the high directivity case, at least one element needs to be aligned.
The more a user is received towards the end-fire direction of the \gls{ula}, the better is the half circle configuration in comparison.
Eventually, the half circle \gls{bs} provides a more evenly distributed coverage and is less sensitive to the direction of the incoming waves.
Furthermore, in multi-user applications, different users are more likely to have stronger contributions to different antenna element subsets of the half circle array.

\begin{figure*}
\begin{subfigure}[b]{.5\linewidth}
\input{tikz/bs_comp_ld_0.tikz}
\caption{$\mathcal{K}=0$}
\end{subfigure}%
\begin{subfigure}[b]{.5\linewidth}
\input{tikz/bs_comp_ld_4.tikz}
\caption{$\mathcal{K}=4$}
\end{subfigure}
\caption{Local diversities are shown for the \gls{ula} (solid line) and half circle (dashed line) \gls{bs} with low directivity elements ($\zeta=2$). The different colours correspond to different \gls{pas}. In general, the half circle \gls{bs} has less local diversity at a certain outage probability. Furthermore, a higher $\mathcal{K}$-factor gives higher local diversity. All constricted cases of diffuse scattering provide less local diversity then the uncorrelated Rayleigh (solid black) or Rice channel (dotted black).} \label{fig:bs_lds}
\end{figure*}
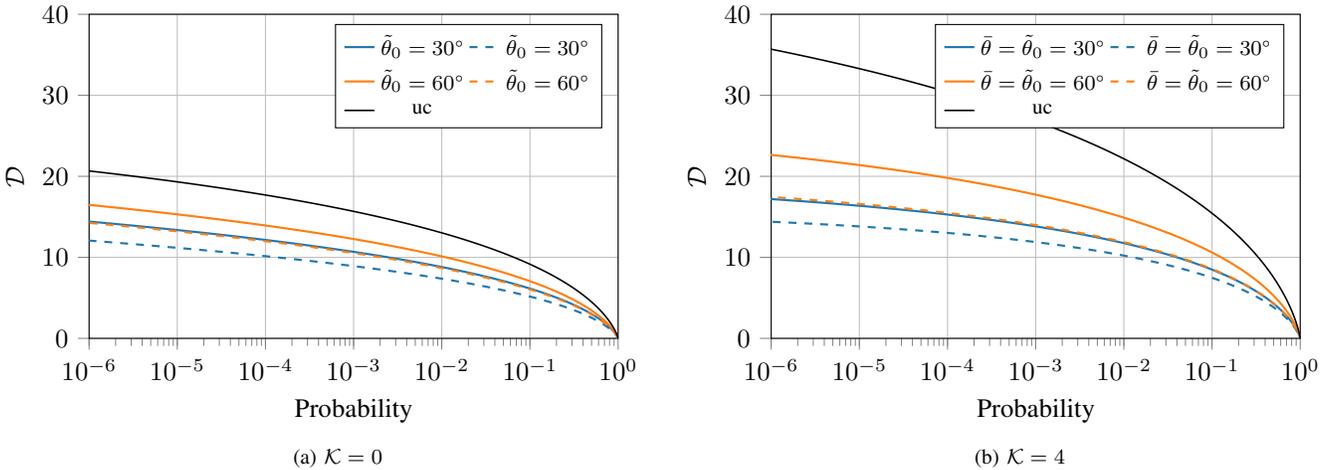

Fig. \ref{fig:bs_lds} presents the local diversities for the two \gls{bs} under different fading conditions.
Only the low directivity case ($\zeta=2$) has been taken into account, to allow for visual comparison of the influence of different \glspl{pas}.
Both, uncorrelated Rayleigh and Rician fading provide more local diversity than their counterparts with constricted \gls{pas}.
The half circle \gls{bs} has always less local diversity in comparison to the \gls{ula} \gls{bs}.
This is due to fewer antenna elements being illuminated by the diffuse component, that provides spatial diversity.
A higher $\mathcal{K}$-factor provides a higher local diversity for all cases, since it becomes more unlikely that the diffuse component of the channel cancels the deterministic component out. 

%% file: tikz/verification_cdf.tikz
\begin{tikzpicture}
\begin{axis}[
legend pos=north east,
legend columns=1, 
tick align=outside,
tick pos=left,
xlabel={$\mathcal{Q}$ [\si{\deci\bel}]},
xmajorgrids,
xmin=-3, xmax=3,
ylabel={CDF},
ymode=log,
ymajorgrids,
ymin=1e-4, ymax=1,
width=.95\linewidth, height=.65\linewidth,
]

\addplot [thick, color0] table[x index=0, y index=1] {data/a1_verify_uc-K-4.0.csv};
\addlegendentry{\footnotesize uc CDF};
\addplot [only marks, color0, mark=x, mark repeat=20, mark phase=7] table[x index=0, y index=2] {data/a1_verify_uc-K-4.0.csv};
\addlegendentry{\footnotesize uc ECDF};

\addplot [thick, color1] table[x index=0, y index=1] {data/a1_verify_omni-K-4.0.csv};
\addlegendentry{\footnotesize omni CDF};
\addplot [only marks, color1, mark=x, mark repeat=20, mark phase=7] table[x index=0, y index=2] {data/a1_verify_omni-K-4.0.csv};
\addlegendentry{\footnotesize omni ECDF};

\addplot [thick, color2] table[x index=0, y index=1] {data/a1_verify_vM_aligned-K-4.0.csv};
\addlegendentry{\footnotesize vM a CDF};
\addplot [only marks, color2, mark=x, mark repeat=20, mark phase=7] table[x index=0, y index=2] {data/a1_verify_vM_aligned-K-4.0.csv};
\addlegendentry{\footnotesize vM a ECDF};

\addplot [thick, color3] table[x index=0, y index=1] {data/a1_verify_vM_squint-K-4.0.csv};
\addlegendentry{\footnotesize vM s CDF};
\addplot [only marks, color3, mark=x, mark repeat=20, mark phase=7] table[x index=0, y index=2] {data/a1_verify_vM_squint-K-4.0.csv};
\addlegendentry{\footnotesize vM s ECDF};

\end{axis}
\end{tikzpicture}

%% file: tikz/verification_ld.tikz
\begin{tikzpicture}
\begin{axis}[
legend pos=north east,
legend columns=1, 
tick align=outside,
tick pos=left,
xlabel={Probability},
xmajorgrids,
xmode=log,
xmin=1e-4, xmax=1,
ylabel={$\mathcal{D}$},
ymajorgrids,
ymin=1, ymax=40,
width=.95\linewidth, height=.65\linewidth,
]

\addplot [thick, color0] table[x index=1, y index=3] {data/a1_verify_uc-K-4.0.csv};
\addlegendentry{\footnotesize uc LD};

\addplot [thick, color1] table[x index=1, y index=3] {data/a1_verify_omni-K-4.0.csv};
\addlegendentry{\footnotesize omni LD};

\addplot [thick, color2] table[x index=1, y index=3] {data/a1_verify_vM_aligned-K-4.0.csv};
\addlegendentry{\footnotesize vM a LD};

\addplot [thick, color3] table[x index=1, y index=3] {data/a1_verify_vM_squint-K-4.0.csv};
\addlegendentry{\footnotesize vM s LD};

\end{axis}
\end{tikzpicture}

%% file: tikz/bs_comp_0_60.tikz
\begin{tikzpicture}
\begin{axis}[
legend pos=north west,
legend columns=1, 
tick align=outside,
tick pos=left,
xlabel={$\mathcal{Q}$ [\si{\deci\bel}]},
xmajorgrids,
xmin=-10, xmax=4,
ylabel={CDF},
ymode=log,
ymajorgrids,
ymin=1e-4, ymax=1,
width=.95\linewidth, height=.65\linewidth,
]

\addplot [thick, color0] table[x index=0, y index=1] {data/a2_ula_vM_aligned-K-0.0-n-2-inc-60.csv};
\addlegendentry{\footnotesize $\zeta=2$};

\addplot [thick, color1] table[x index=0, y index=1] {data/a2_ula_vM_aligned-K-0.0-n-5-inc-60.csv};
\addlegendentry{\footnotesize $\zeta=5$};

\addplot [thick, color2] table[x index=0, y index=1] {data/a2_ula_vM_aligned-K-0.0-n-20-inc-60.csv};
\addlegendentry{\footnotesize $\zeta=20$};

\addplot [thick, color0, dashed] table[x index=0, y index=1] {data/a4_circ_vM_aligned-K-0.0-n-2-inc-60.csv};
\addlegendentry{\footnotesize $\zeta=2$};

\addplot [thick, color1, dashed] table[x index=0, y index=1] {data/a4_circ_vM_aligned-K-0.0-n-5-inc-60.csv};
\addlegendentry{\footnotesize $\zeta=5$};

\addplot [thick, color2, dashed] table[x index=0, y index=1] {data/a4_circ_vM_aligned-K-0.0-n-20-inc-60.csv};
\addlegendentry{\footnotesize $\zeta=20$};

\end{axis}
\end{tikzpicture}

%% file: tikz/bs_comp_4_60.tikz
\begin{tikzpicture}
\begin{axis}[
legend pos=north west,
legend columns=1, 
tick align=outside,
tick pos=left,
xlabel={$\mathcal{Q}$ [\si{\deci\bel}]},
xmajorgrids,
xmin=-10, xmax=4,
ylabel={CDF},
ymode=log,
ymajorgrids,
ymin=1e-4, ymax=1,
width=.95\linewidth, height=.65\linewidth,
]

\addplot [thick, color0] table[x index=0, y index=1] {data/a2_ula_vM_aligned-K-4.0-n-2-inc-60.csv};
\addlegendentry{\footnotesize $\zeta=2$};

\addplot [thick, color1] table[x index=0, y index=1] {data/a2_ula_vM_aligned-K-4.0-n-5-inc-60.csv};
\addlegendentry{\footnotesize $\zeta=5$};

\addplot [thick, color2] table[x index=0, y index=1] {data/a2_ula_vM_aligned-K-4.0-n-20-inc-60.csv};
\addlegendentry{\footnotesize $\zeta=20$};

\addplot [thick, color0, dashed] table[x index=0, y index=1] {data/a4_circ_vM_aligned-K-4.0-n-2-inc-60.csv};
\addlegendentry{\footnotesize $\zeta=2$};

\addplot [thick, color1, dashed] table[x index=0, y index=1] {data/a4_circ_vM_aligned-K-4.0-n-5-inc-60.csv};
\addlegendentry{\footnotesize $\zeta=5$};

\addplot [thick, color2, dashed] table[x index=0, y index=1] {data/a4_circ_vM_aligned-K-4.0-n-20-inc-60.csv};
\addlegendentry{\footnotesize $\zeta=20$};

\end{axis}
\end{tikzpicture}

%% file: tikz/bs_comp_0_30.tikz
\begin{tikzpicture}
\begin{axis}[
legend pos=south east,
legend columns=1, 
tick align=outside,
tick pos=left,
xlabel={$\mathcal{Q}$ [\si{\deci\bel}]},
xmajorgrids,
xmin=-10, xmax=4,
ylabel={CDF},
ymode=log,
ymajorgrids,
ymin=1e-4, ymax=1,
width=.95\linewidth, height=.65\linewidth,
]

\addplot [thick, color0] table[x index=0, y index=1] {data/a2_ula_vM_aligned-K-0.0-n-2-inc-30.csv};
\addlegendentry{\footnotesize $\zeta=2$};

\addplot [thick, color1] table[x index=0, y index=1] {data/a2_ula_vM_aligned-K-0.0-n-5-inc-30.csv};
\addlegendentry{\footnotesize $\zeta=5$};

\addplot [thick, color2] table[x index=0, y index=1] {data/a2_ula_vM_aligned-K-0.0-n-20-inc-30.csv};
\addlegendentry{\footnotesize $\zeta=20$};

\addplot [thick, color0, dashed] table[x index=0, y index=1] {data/a4_circ_vM_aligned-K-0.0-n-2-inc-30.csv};
\addlegendentry{\footnotesize $\zeta=2$};

\addplot [thick, color1, dashed] table[x index=0, y index=1] {data/a4_circ_vM_aligned-K-0.0-n-5-inc-30.csv};
\addlegendentry{\footnotesize $\zeta=5$};

\addplot [thick, color2, dashed] table[x index=0, y index=1] {data/a4_circ_vM_aligned-K-0.0-n-20-inc-30.csv};
\addlegendentry{\footnotesize $\zeta=20$};

\end{axis}
\end{tikzpicture}

%% file: tikz/bs_comp_4_30.tikz
\begin{tikzpicture}
\begin{axis}[
legend pos=south east,
legend columns=1, 
tick align=outside,
tick pos=left,
xlabel={$\mathcal{Q}$ [\si{\deci\bel}]},
xmajorgrids,
xmin=-10, xmax=4,
ylabel={CDF},
ymode=log,
ymajorgrids,
ymin=1e-4, ymax=1,
width=.95\linewidth, height=.65\linewidth,
]

\addplot [thick, color0] table[x index=0, y index=1] {data/a2_ula_vM_aligned-K-4.0-n-2-inc-30.csv};
\addlegendentry{\footnotesize $\zeta=2$};

\addplot [thick, color1] table[x index=0, y index=1] {data/a2_ula_vM_aligned-K-4.0-n-5-inc-30.csv};
\addlegendentry{\footnotesize $\zeta=5$};

\addplot [thick, color2] table[x index=0, y index=1] {data/a2_ula_vM_aligned-K-4.0-n-20-inc-30.csv};
\addlegendentry{\footnotesize $\zeta=20$};

\addplot [thick, color0, dashed] table[x index=0, y index=1] {data/a4_circ_vM_aligned-K-4.0-n-2-inc-30.csv};
\addlegendentry{\footnotesize $\zeta=2$};

\addplot [thick, color1, dashed] table[x index=0, y index=1] {data/a4_circ_vM_aligned-K-4.0-n-5-inc-30.csv};
\addlegendentry{\footnotesize $\zeta=5$};

\addplot [thick, color2, dashed] table[x index=0, y index=1] {data/a4_circ_vM_aligned-K-4.0-n-20-inc-30.csv};
\addlegendentry{\footnotesize $\zeta=20$};

\end{axis}
\end{tikzpicture}

%% file: tikz/bs_comp_ld_0.tikz
\begin{tikzpicture}
\begin{axis}[
legend pos=north east,
legend columns=2, 
tick align=outside,
tick pos=left,
xlabel={Probability},
xmajorgrids,
xmode=log,
xmin=1e-6, xmax=1,
ylabel={$\mathcal{D}$},
ymajorgrids,
ymin=0, ymax=40,
width=.95\linewidth, height=.65\linewidth,
]

\addplot [thick, color0] table[x index=1, y index=2] {data/a2_ula_vM_aligned-K-0.0-n-2-inc-30.csv};
\addlegendentry{\footnotesize $\tilde\theta_0 = \ang{30}$};

\addplot [thick, color0, dashed] table[x index=1, y index=2] {data/a4_circ_vM_aligned-K-0.0-n-2-inc-30.csv};
\addlegendentry{\footnotesize $\tilde\theta_0 = \ang{30}$};

\addplot [thick, color1] table[x index=1, y index=2] {data/a2_ula_vM_aligned-K-0.0-n-2-inc-60.csv};
\addlegendentry{\footnotesize $\tilde\theta_0 = \ang{60}$};

\addplot [thick, color1, dashed] table[x index=1, y index=2] {data/a4_circ_vM_aligned-K-0.0-n-2-inc-60.csv};
\addlegendentry{\footnotesize $\tilde\theta_0 = \ang{60}$};

\addplot [semithick, black] table[x index=2, y index=3] {data/K_-1000-M_32.csv};
\addlegendentry{\footnotesize uc};

\end{axis}
\end{tikzpicture}

%% file: tikz/bs_comp_ld_4.tikz
\begin{tikzpicture}
\begin{axis}[
legend pos=north east,
legend columns=2, 
tick align=outside,
tick pos=left,
xlabel={Probability},
xmajorgrids,
xmode=log,
xmin=1e-6, xmax=1,
ylabel={$\mathcal{D}$},
ymajorgrids,
ymin=0, ymax=40,
width=.95\linewidth, height=.65\linewidth,
]

\addplot [thick, color0] table[x index=1, y index=2] {data/a2_ula_vM_aligned-K-4.0-n-2-inc-30.csv};
\addlegendentry{\footnotesize $\bar\theta = \tilde\theta_0 = \ang{30}$};

\addplot [thick, color0, dashed] table[x index=1, y index=2] {data/a4_circ_vM_aligned-K-4.0-n-2-inc-30.csv};
\addlegendentry{\footnotesize $\bar\theta = \tilde\theta_0 = \ang{30}$};

\addplot [thick, color1] table[x index=1, y index=2] {data/a2_ula_vM_aligned-K-4.0-n-2-inc-60.csv};
\addlegendentry{\footnotesize $\bar\theta = \tilde\theta_0 = \ang{60}$};

\addplot [thick, color1, dashed] table[x index=1, y index=2] {data/a4_circ_vM_aligned-K-4.0-n-2-inc-60.csv};
\addlegendentry{\footnotesize $\bar\theta = \tilde\theta_0 = \ang{60}$};

\addplot [semithick, black] table[x index=2, y index=3] {data/K_6-M_32.csv}; %
\addlegendentry{\footnotesize uc};

\end{axis}
\end{tikzpicture}

%% file: discussion.tex
\section{Discussion}

The proposed correlated Rician model increases the realism and complexity of a massive \gls{mimo} \gls{bs} propagation analysis over simpler uncorrelated models.
Additionally, antenna array geometry, antenna element orientation and \glspl{pas} have been accounted for to determine the covariance matrix of the \gls{cnrv}, allowing for evaluation of different \gls{bs} designs for a given scenario.
The parameters of the \gls{cnrv} can alternatively be based on the simplified \gls{3gpp} Urban Microcell model as described in \cite{polegre_channel_2020} and our \glspl{cgqf} approach still gives the \gls{pdf} and \gls{cdf} of the effective channel.
Furthermore, the method can be generalised to handle e.g. power variation between antenna elements through the operator matrix. 

Covariance is sometimes elusive and hard to measure properly in the field.
However, by inspection of the quadratic form in \eqref{eqn:qf_eigen}, it is clear, that the properties of the eigenvalues of the covariance matrix are of central relevance to capture the effect of correlation on the effective channel.
Moreover, considering \emph{universality} in \gls{rmt} \cite{edelman_beyond_2016}, the distribution of eigenvalues behaves asymptotically as if the matrix elements are Gaussian distributed.
Therefore, the \gls{cgqf} results are less sensitive to the actual individual correlation coefficients. 

So far, we only demonstrated over-the-horizon propagation with the example \glspl{pas}.
Nonetheless, the described framework allows for three-dimensional considerations.
The impact of incoming waves from different elevation angles depends obviously on the weighting imposed by antenna element pattern and array geometry.
In general, additional local diversity is available, but the \gls{bs} design needs to take this into account to benefit from it.
The proposed model is flexible enough to allow \gls{bs} performance comparisons for a combination of surface and aerial users. 

%% file: conclusion.tex
\section{Conclusion}
A thorough way of handling correlated Rician fading for \glspl{lsas} has been presented.
The accurate approximations can be used to analyse the effective channel of massive \gls{mimo} \glspl{bs}.
The framework allows consideration of inter-tap correlation in addition to the outlined antenna correlations.
Inter-tap correlation can occur, e.g. if distributed antenna arrays are spaced further apart than the distance related to the duration of a single tap.

The provided correlated \gls{cnrv} channel model is general and the $\mathcal{K}$-factor parameterisation allows investigation of Rician and Rayleigh fading at each antenna and delay tap with arbitrary correlation coefficients.
Correlation coefficients have been related to the \gls{pas} of the diffuse channel component and consider the antenna element pattern in additon.
A plane wave model provides the foundation of the \gls{cn} element statistics and the tractability of correlation coefficients.

Low-directivity antenna elements provide better overall system performance than high-directivity elements, once the \gls{bs} grows to a reasonable size.
This is mainly caused by being less prone to the direction of incoming waves, because the main gain is coming from the steerable array factor and not the static element factor.
Arranging the elements in a half-circle illuminates a region more evenly than a \gls{ula}, but reduces the peak gain for broad-side radiation.
Moreover, the local diversity is reduced for the half-circle \gls{bs} and a directional diffuse part since fewer elements pick up significant energy, even though they are less correlated.

The complete statistic of the effective channel gain for correlated Rician fading channels is described through the provided \gls{pdf} or \gls{cdf}.
This allows the investigation of instantaneous metrics of the single user performance beyond the mean and variance of the combined received signal.